\documentclass[sigconf]{acmart}

\AtBeginDocument{%
  \providecommand\BibTeX{{%
    \normalfont B\kern-0.5em{\scshape i\kern-0.25em b}\kern-0.8em\TeX}}}

\copyrightyear{2020} 
\acmYear{2020} 
\setcopyright{none}
\acmConference[EASE 2020]{Evaluation and Assessment in Software Engineering}{April 15--17, 2020}{Trondheim, Norway}
\acmBooktitle{Evaluation and Assessment in Software Engineering (EASE 2020), April 15--17, 2020, Trondheim, Norway}
\acmPrice{}
\acmDOI{}
\acmISBN{}

\usepackage{adjustbox}
\usepackage{listings}
\usepackage{color, colortbl}
\usepackage{multirow}

\usepackage{xspace}
\usepackage{graphicx}
\usepackage{ifthen}
\usepackage[normalem]{ulem} 
\usepackage{xcolor,colortbl}
\usepackage{hyperref}

\newboolean{showedits}
\setboolean{showedits}{true} 
\ifthenelse{\boolean{showedits}}
{
	\newcommand{\del}[1]{\textcolor{red}{\sout{#1}}} 
	\newcommand{\nbe}[3]{
		{\colorbox{#3}{\bfseries\sffamily\scriptsize\textcolor{white}{#1}}}
		{\textcolor{#3}{\sf\small$\blacktriangleright$\textit{#2}$\blacktriangleleft$}}}
}{
	\newcommand{\del}[1]{} 
	
	\newcommand{\nbe}[3]{}
}

\newboolean{showcomments}
\setboolean{showcomments}{true} 
\newcommand{\id}[1]{$-$Id: scgPaper.tex 32478 2010-04-29 09:11:32Z oscar $-$}

\ifthenelse{\boolean{showcomments}}
 {
 	\newcommand{\nbc}[3]{
 		{\colorbox{#3}{\bfseries\sffamily\scriptsize\textcolor{white}{#1}}}
		{\textcolor{#3}{\sf\small$\blacktriangleright$\textit{#2}$\blacktriangleleft$}}}
	
 }{
 	\newcommand{\nbc}[3]{}
 	
 }

\usepackage[most]{tcolorbox}
\ifthenelse{\boolean{showedits}}
{
  \newtcolorbox{inserted}{%
       title=Inserted text:,
       colframe=blue,colback=blue!5!white,
       breakable,
       leftrule=0mm, 
       bottomrule=0mm,
       rightrule=0mm,
       toprule=0mm,
       arc=0mm, outer arc=0mm,
       oversize
  }
  \newtcolorbox{deleted}{%
       title=Deleted text:,
       colframe=red,colback=red!5!white,
       breakable,
       leftrule=0mm, 
       bottomrule=0mm,
       rightrule=0mm,
       toprule=0mm,
       arc=0mm, outer arc=0mm,
       oversize
  }
  \newtcolorbox{refactored}{%
       title=Rewritten text:,
       colframe=blue,colback=red!5!white,
       breakable,
       leftrule=0mm, 
       bottomrule=0mm,
       rightrule=0mm,
       toprule=0mm,
       arc=0mm, outer arc=0mm,
       oversize
  }
}{

}
\newboolean{isblinded}
\setboolean{isblinded}{true}
\ifthenelse{\boolean{isblinded}}
{\newcommand\blind[1]{BLINDED\xspace}}
{\newcommand\blind[1]{#1\xspace}}

\newcommand{\commented}[1]{}

\newcommand{\eg}{\emph{e.g.,}\xspace}
\newcommand{\ie}{\emph{i.e.,}\xspace}
\newcommand{\etal}{\emph{et al.}\xspace}

\lstdefinelanguage{Java}{
  tabsize=4
}[keywords,comments,strings]

\definecolor{source}{gray}{0.95}
\definecolor{highlight}{gray}{0.9}

\lstnewenvironment{codesnippet}{%
	\lstset{%
		frame=single,
		framerule=0pt,
		mathescape=false
	}
}{}

\makeatletter                  
\def\mdseries@tt{m}      
\makeatother                   

\begin{document}

\title{Tricking Johnny into Granting Web Permissions}



\author{Mohammadreza Hazhirpasand}
\email{mohammadreza.hazhirpasand@inf.unibe.ch}
\affiliation{%
  \institution{University of Bern}
}
\author{Mohammad Ghafari}
\email{mohammad.ghafari@inf.unibe.ch}
\orcid{0000-0002-1986-9668}
\affiliation{%
  \institution{University of Bern}
}
\author{Oscar Nierstrasz}
\email{oscar.nierstrasz@inf.unibe.ch}
\orcid{0000-0002-9975-9791}
\affiliation{%
  \institution{University of Bern}
}
\author{} 

\begin{abstract}
We studied the web permission API dialog box in popular mobile and desktop browsers, and found that it typically lacks measures to protect users from unwittingly granting web permission when clicking too fast.

We developed a game that exploits this issue, and tricks users into granting webcam permission.
We conducted three experiments, each with 40 different participants, on both desktop and mobile browsers.
The results indicate that in the absence of a prevention mechanism,
we achieve a considerably high success rate in tricking 95\% and 72\% of participants on mobile and desktop browsers, respectively.
Interestingly, we also tricked 47\% of participants on a desktop browser where a prevention mechanism exists.

\end{abstract}



\maketitle

\section{Introduction}
\label{sec:intro}

In a clickjacking attack, an attacker tricks a user into clicking on webpage element that is hidden or disguised as another element.
This can result in a situation in which the user takes an action unwittingly.
For instance, attackers have used clickjacking attacks to trick users into liking a fan page on Facebook or re-tweeting a message on Twitter \cite{balduzzi2010solution}.
Several clickjacking attacks targeted the Adobe Flash Player's webcam access dialog to enable the victims' webcam and microphone \cite{huang2012clickjacking}.
Adobe finally fixed the issue by ensuring that the webcam access dialog is fully visible to users.

There are a number of ways in which clickjacking attacks can be prevented. The widely accepted approach is to use \emph{frame busting} techniques.
Frame busting refers to client-side code that is designed to prevent a given web page from being loaded in a sub-frame.
Many JavaScript code snippets have been proposed to perform frame busting, though many of these have been found to be vulnerable later \cite{rydstedt2010busting}.
A more robust solution is to send a special HTTP header, \eg X-Frame-Options, which is supported by some browsers such as Mozilla Firefox and Google Chrome \cite{niemietz2011ui}.
The X-Frame-Options header prohibits a website from being rendered within an iframe.
However, some older versions of browsers do not support the special header that prevents Clickjacking attacks.
Buchanan \etal analyzed Alexa's top one million sites and their results show that X-Frame-Options is implemented in only a small fraction (\ie 11.11\%) of the websites \cite{buchanan2017analysis}.

In this paper, we discuss the lack of a preventive measure, \eg a delay period, in many browsers when the web permission dialog box pops up.
This is a browser-dependent feature, and none of the aforementioned countermeasures could guarantee user safety.
In order to highlight the importance of this issue, we investigate the following research question:
\emph{``How can the web permission dialog box be abused, and how effective are the existing preventive measures in browsers?''}
To this end, we designed an experimental game, called ``Furious Clicker,'' to evaluate whether or not users can be tricked into granting webcam permission by clicking on the \emph{allow} button of the web permission API's dialog box.
We asked 120 participants to take part in our web-based game experiment.
We conducted three experiments, each with 40 different participants, on both desktop and mobile browsers.
We used the mobile version of Google Chrome on Android, and the desktop version of Mozilla Firefox and Google Chrome on Mac OS.
The results indicate that, in the absence of a prevention mechanism, we achieve a considerably high success rate in tricking 95\% of participants on a mobile browser, and 72\% on a desktop browser.
We also tricked 47\% of participants on a desktop browser where a prevention mechanism exists.

We conclude that without a preventive measure end users suffer from severe security implications. 
For instance, the Android version of the Google Chrome browser, with more than five billion downloads on Google Play, has no prevention mechanism.
Nevertheless, the impact of any countermeasure, such as a delay when the permission dialog box appears, on the user experience needs to be taken into account.

\begin{table*}[]
\centering
\caption{Tested browsers and position of permission dialog box
}
\label{tbl:browsers}
\begin{tabular}{|l|l|l|l|l|l|l|}
\hline
\textbf{Browser}                                    & \textbf{Version} &\textbf{Platform}               & \textbf{Position} & \textbf{Alert}        & \textbf{Behavior} & \textbf{Prevention} \\ \hline
Firefox & 68      & macOS - desktop  & Top Left         & Icon  & Ask again    & Yes        \\ \hline
Chrome   & 77      & macOS - desktop  & Top Left         & Icon  & Never ask    & Yes        \\ \hline
Safari                                           & 13      & macOS - desktop  & Top Center       & Icon  & Ask again    & No         \\ \hline
Chrome                                    & 78      & Android - mobile  & Center           & Notification & Never ask    & No         \\ \hline
Firefox                                  & 68      & Android - mobile  & Top              & Notification & Ask again    & No         \\ \hline
Edge                                             & 42      & Android - mobile  & Center           & Notification & Never ask    & No         \\ \hline
Dolphin                                  & 8       & Android - mobile  & Center           & -            & Never ask    & No         \\ \hline
Mint                                     & 61      & Android - mobile  & Bottom           & -            & Never ask    & No         \\ \hline
Samsung                                  & 10      & Android - mobile  & Bottom           & -            & Never ask    & No         \\ \hline
Mi Browser                                 & 11     & Android - mobile  & Bottom           & -           & Never ask   & No         \\ \hline
UC Browser                                 & 12     & Android - mobile  & Bottom           & -           & Never ask   & No         \\ \hline
Edge                                  & 42      & Windows - desktop  & Bottom           & Icon          & Ask again    & No         \\ \hline
Konqueror                                  & 5      & Linux - desktop  & Center           & -           & Ask again    & No         \\ \hline
Web                                  & 3     & Linux - desktop  & Top           & -           & Never ask   & No         \\ \hline
\end{tabular}
\end{table*}

The remainder of this paper is structured as follows.
In \autoref{sec:clickattacks}, we explain our motivation and various types of clickjacking attacks.
In \autoref{sec:ce} we introduce the Furious Clicker game.
In \autoref{sec:study} we present a user experiment study, and discuss related work in \autoref{sec:related}.
We conclude this paper in \autoref{sec:conclusion}.

\section{Motivation}
\label{sec:clickattacks}

We present the web permission API, how it can be abused, and what factors are essential to consider in abusing the web permission API.

\subsection{The web permission API}
The web permission API offers a uniform way for websites to ask users' permission for critical features that require user consent, such as camera, clipboard, microphone, or geolocation.
When a website requires a specific permission, a dialog box appears in the browser and the user can grant, deny, or dismiss the permission.
Users can also revoke or grant the permission later in the browsers' settings.
Nevertheless, for inexperienced users, it is challenging to find out where the permissions, which have been asked previously,  are in browsers' settings.

The permission API behaves differently in different browsers.
For example, a website cannot ask twice for a denied permission on the desktop version of Google Chrome, while it can do so after reloading on the desktop version of Mozilla Firefox.

\subsection{Clickjacking attacks}
There exist three main types of clickjacking attacks.
In the first attack, \emph{Jeopardizing target display integrity}, adversaries attempt to either set a sensitive UI element to be transparent and place an attractive decoy element beneath the main element, or they cover part of a sensitive UI element by overlaying a decoy element \cite{stone2010next}.\footnote{https://feross.org/webcam-spy}

In the second attack, \emph{Jeopardizing pointer integrity}, adversaries show a fake cursor (pointer) instead of the real cursor to victims, and conceal the default cursor programmatically \cite{bordi2010proof} \cite{shahriar2015request}. 

In contrast to the two previous attack types, in the \emph{Jeopardizing temporal integrity} attack, adversaries set a sensitive element to appear on top of a decoy element when users are busy clicking on the decoy element.
The users observe the sensitive element, but are trapped into making an unwanted click due to having a very limited time to stop an action\,---\,humans need at least a few hundred milliseconds to react to a sudden visual change.
Adversaries use this type of attacks in online games, \ie Adobe Flash webcam access, and security dialog bypass by captcha.\footnote{https://www.squarefree.com/2004/07/01/race-conditions-in-security-dialogs/} \footnote{https://bugzilla.mozilla.org/show\_bug.cgi?id=162020}

\subsection{The problem}
\label{sec:theprob}
The usual way to keep users safe from \emph{temporal integrity} attacks is to provide them with enough time to grasp any UI changes.
For instance, to install a Chrome extension, users need to wait until the delay expires.

We found that almost none of the popular browsers take such a preventive measure when presenting the web permission API dialog box.
The dialog box often appears in the area where websites present some content, \eg images or clickable elements, and therefore users might click on the \emph{allow} button inadvertently.

We tested 14 browsers to collect data about where the permission dialog box appears, how each browser informs users that the webcam or microphone is in use, and how each browser behaves
when the web permission dialog box is dismissed.
The position of the dialog box depends on several factors.
The first factor is that the buttons of the dialog box have different sizes in different browsers.
For instance, in Safari, the buttons are narrower compared to those of Chrome, where the buttons are rectangular.
Another factor, which is user-dependent, is that
different types of bars below the address bar in browsers affect the position of the permission dialog box.
For instance, the bookmark bar on the Chrome browser makes the dialog box appear at different coordinates.
\autoref{tbl:browsers} shows the results of our observations.
For instance, in the desktop version of Safari, version 13,  the web permission dialog box appears in the top center of screen and informs users that the webcam is in use by placing a camera icon on the website's tab.
In case a user clicks on the \emph{deny} button of specific permission, an attacker can trigger the same permission under the same domain name more than once.
In desktop browsers, the dialog box becomes visible usually on the top part of the browsers under the address bar.
In all tested desktop browsers, there is a small blinking icon next to the website's icon, intended to catch the attention of users.
In contrast, mobile browsers show a notification in the notification bar which clearly notifies users that the website is using your camera or microphone.
Firefox, Konqueror, and Safari behave in such a way that websites can request specific permission more than once when users dismiss the permission for the first time.
In contrast, the other browsers do not allow websites to ask for the same permission after being dismissed by users.
Therefore, adversaries need to be careful in drafting their attack scenarios to precisely land a user's click on the \emph{allow} button of the dialog box for such browsers.
Only the desktop version of Firefox and Chrome accept user clicks with a short delay when a user clicks rapidly before the permission dialog box appears.
The other tested browsers do not offer any preventive measures at the time of writing.

\begin{table*}[]
\centering
\caption{The three experiments for evaluating Furious Clicker }
\label{tbl:exp}
\begin{tabular}{|l|l|l|l|l|l|}
\hline
\textbf{Browser} & \textbf{Version} & \textbf{Platform}  & \textbf{Device} & \textbf{Prevention}  & \textbf{\# of Participants} \\ \hline
Chrome        & 72      & MacOS & Mouse           & Yes - delay        & 40                 \\ \hline
Safari        & 13      & MacOS  & Mouse           & No                 & 40                 \\ \hline
Chrome        & 78      & Android & Finger touch    & No                 & 40                 \\ \hline
\end{tabular}
\end{table*}

\section{Furious Clicker}
\label{sec:ce}

We now describe how we designed a game, called \emph{Furious Clicker},\footnote{https://www.crypto-explorer.com/clickjacking} to land a user click on the \emph{allow} button of the web permission dialog box.

\begin{figure}
\includegraphics[width=1\linewidth]{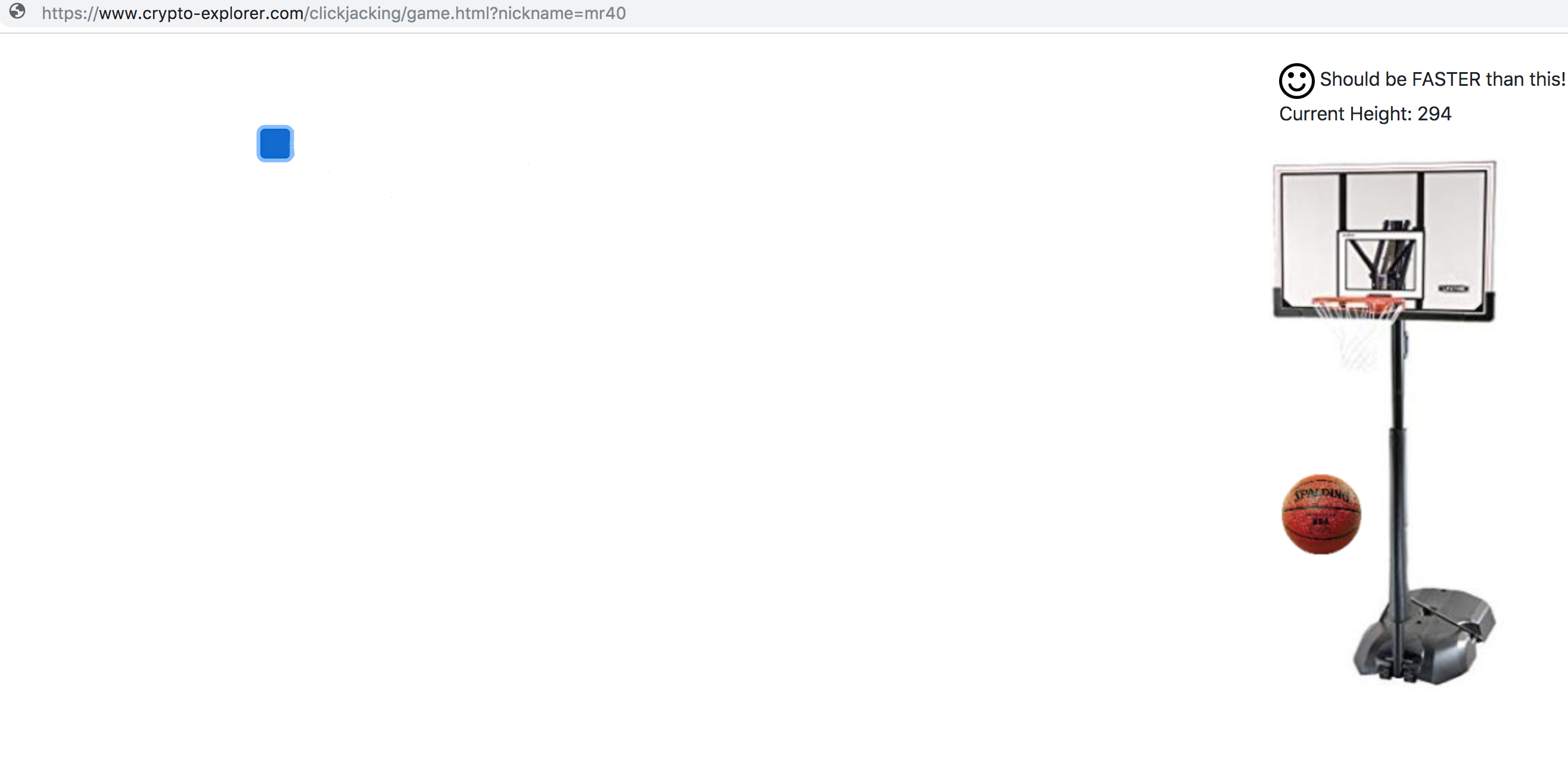}
\caption{The desktop version of Furious Clicker}\label{fig:mobile}
\end{figure}

The game is designed in such a way as to persuade users to click quickly on an HTML element in order to complete an engaging task.
The goal of Furious Clicker is to lift a basketball up and eventually drop it into a net, as shown in \autoref{fig:mobile}.
In the game, however, gravity is strong.
To lift the basketball up, users need to defeat gravity by clicking very fast on a blue button, otherwise the ball will not go up far enough to fall into the net.
The faster a user clicks on the button, the more the basketball goes up.
We measure the time between two clicks of a user.
If the time is very short (\eg less than 100 milliseconds), it conveys that the user is clicking fast and we raise the basketball more quickly.

\begin{figure}
\includegraphics[width=1\linewidth]{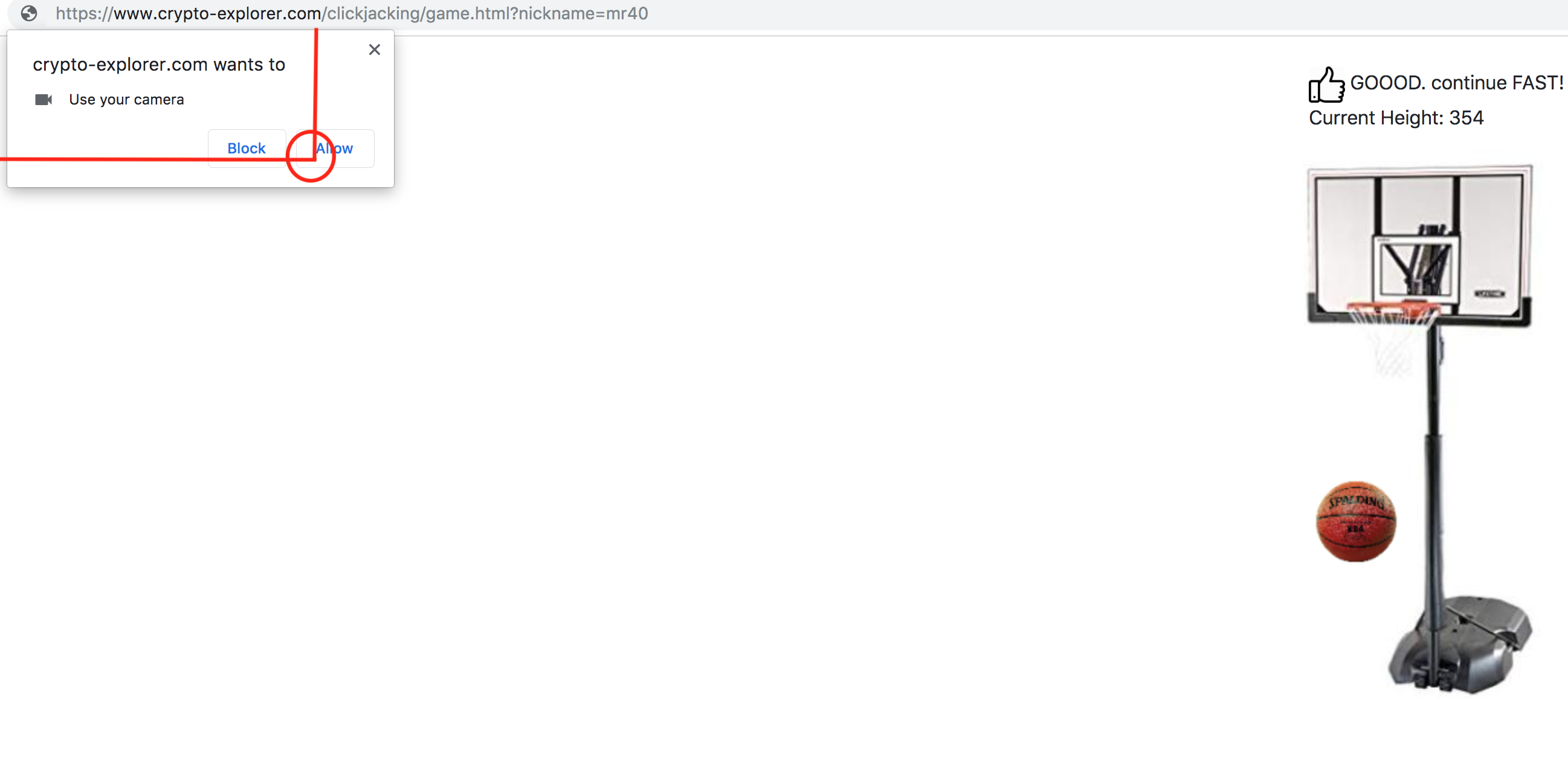}
\caption{The web permission dialog box's position in the game}\label{fig:desktop}
\end{figure}

In the beginning, the game asks the user to input a nickname, which is combined with a random number.
In case the user clicks at a slow pace, the game encourages the user to click faster by displaying a message.
The game issues two more messages depending on user's clicking speed, to persuade them to reach the highest speed.
Eventually, when the ball reaches the net, the game displays a congratulatory message for completing the game successfully.
On the desktop version of browsers, we place a basketball hoop on the right side of the screen to divert the user's attention to that side.
This cannot be done in mobile browsers as the screen size is too small.
Therefore, touch-based devices require a specific design, such as bigger icons and buttons, as users cannot click accurately on small HTML elements.
The malicious blue button in the game, which is called the decoy element, must be placed in a precise position depending on the victim's user agent and platform, as discussed in \autoref{sec:theprob}, otherwise, the attack will fail (see \autoref{fig:desktop}).

In order to decide when to trigger the web permission dialog box, we established a clicking threshold to estimate the level of user engagement in the game.
We iterated our initial test 20 times with different thresholds for the number of clicks.
We determined that between 8 and 12 consecutive fast clicks indicate a high likelihood that the user will mistakenly click on the \emph{allow} button.
The duration between each click needs to be less than 70 milliseconds, and the number of clicks should be more than 8 consecutive clicks.

Once a user reaches the fast clicking threshold, the game triggers the web permission.
In case the user clicks on the \emph{allow} button, the game takes a photo and silently sends it to the server.
If the user stops clicking and chooses the \emph{deny} button, the game will send a message to the server stating that the player is not tricked.

We examined different color schemes for the background color of Furious Clicker. 
To avoid distracting users, finally, we chose the white background color as it is close to the color of the web permission dialog box.


\section{User Experiment}
\label{sec:study}
We investigate how effective the Furious Clicker is in tricking users to click on the \emph{allow} button when a browser presents a permission  dialog box.

\subsection{Method}
We asked for interested students in twelve classes of a private educational technology and engineering institute to participate in our research study. 
We conducted three experiments with a total of 120 participants who willingly took part in our experiments without being paid.
The participants were all studying practical courses related to engineering or computer science.
They all had an academic background (\ie 67\% Bachelor degree, and 33\% Master degree).
Only 22 had their degree in Computer Science.
None of the participants had knowledge of or expertise in information security.

We designed three different experiments to assess how successful the Furious Clicker is in tricking users to grant the camera permission.
\autoref{tbl:exp} gives an overview of the three experiments.
In particular, we designed one experiment to examine the impact of the existing preventive measure when clicking fast in the desktop version of Chrome; the second experiment tests the Safari browser which offers no preventive measure; and finally, we conduct an experiment with the mobile version of the Chrome browser, which also offers no preventive measures.
In the first two experiments, participants are required to use a mouse, whereas in the last experiment users use the touch feature of the smartphone.

We divided the participants into three equal groups, each consisting of 40 people, and assigned each group to a different experiment.
We first conducted an offline survey to understand the level of familiarity of participants with browsers and web permissions.
We then presented the game to participants of each group and made sure that everyone understands how the game works.
We stated that the goal of the test is to measure how fast people are able to click and evaluates the impact of such games on gamers in which fast clicking is required.
We also observed how each participant engages in the game. 

We interviewed each participant right after playing the game.
We asked their opinion about the game and also what they observed while playing the game.

\subsection{Results}
\begin{figure}
\includegraphics[width=1\linewidth,trim=4 4 4 4,clip]{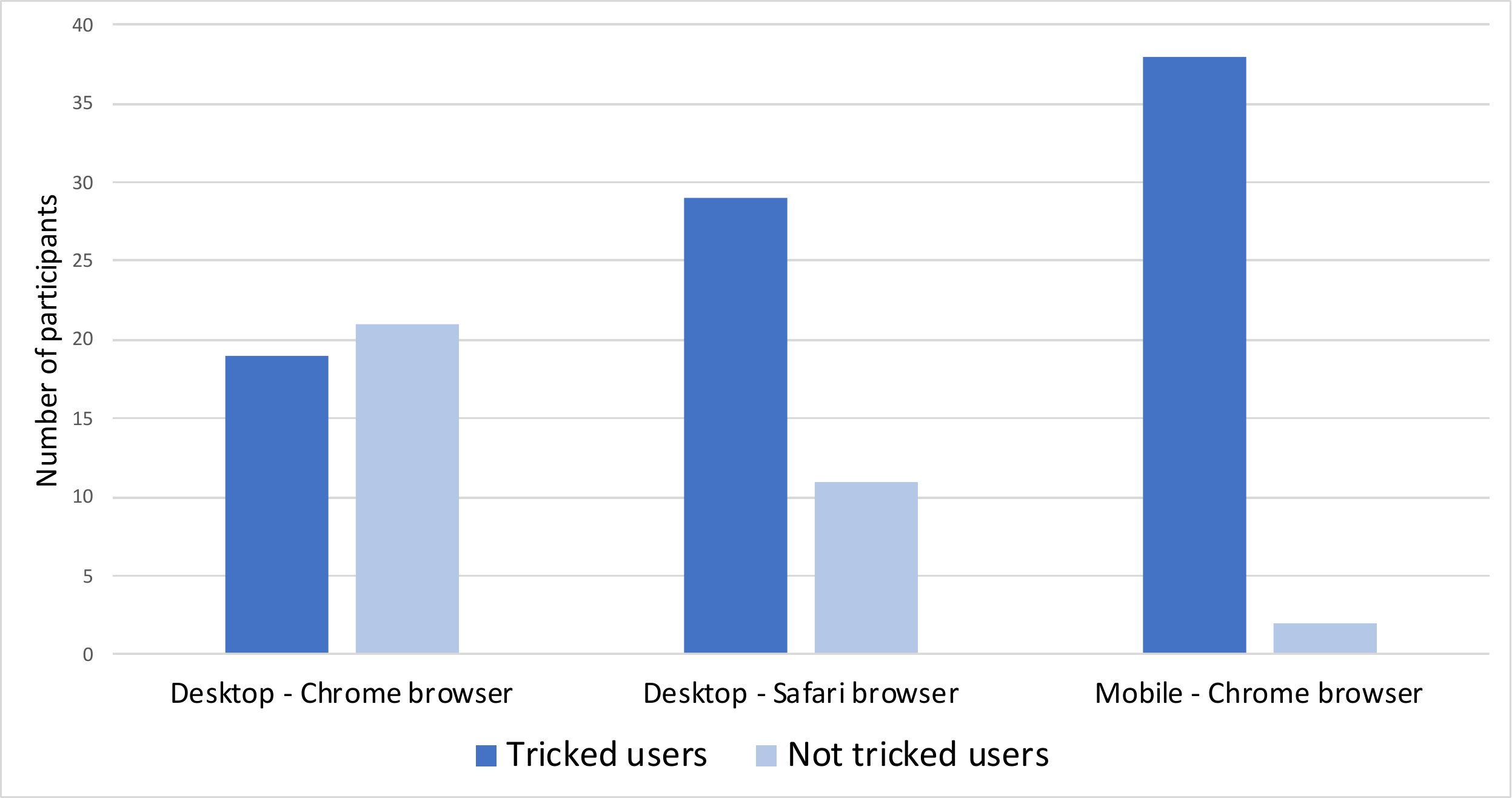}
\caption{The number of participants with different status in each experiment}
\label{fig:experimentres}
\end{figure}

We present the results of the three experiments in \autoref{fig:experimentres}.
In the first experiment, 19 participants clicked on the \emph{allow} button and did not realize what they clicked.
Suspicions of all the remaining 21 participants arose as they observed the web permission dialog asking for camera access.
This is due to the short delay implemented in Google Chrome.
Of the users who were not tricked, 14 stopped playing the game as soon as they saw the permission dialog box pop up, and seven closed the dialog box, and continued playing after a short interruption.

In the second experiment, 29 participants clicked on the \emph{allow} button and finished the game successfully.
Seven contestants were not tricked because of their incorrect mouse coordinates during the fast click process.
Their mouse pointer exited the region of the decoy button when the dialog box appeared.
The rest of the not tricked participants stopped unexpectedly exactly when the game decided to show the web permission dialog box.

In the last experiment, we used the mobile version of Google Chrome on Android devices.
In total, 38 participants were tricked into clicking on the \emph{allow} button and only two of them suddenly stopped while clicking on the button.

We were interested to know how familiar participants are with web permissions. We therefore surveyed participants before conducting the experiments.
To minimize the Hawthorne effect, 
we also populated the survey with some questions regarding other features of browsers, not relevant to web permissions, such as browser history, bookmarks, anonymous mode, and 3D games \cite{mccarney2007hawthorne}.
Almost 83\% of all 120 participants had never played a game on browsers while others had played at least once.
Participants were mainly familiar with the history feature of browsers (95\%).
More than three quarters (83\%) of the participants had their customized bookmarks in their favorite browser.
As the privacy of data has been emerging as a concern for web users, the anonymity feature of browsers had been used by 57\% of the participants.
We also asked if they had any experience in granting permission in the web environment.
Participants did not know much about permissions and how they work in browsers as only 42\% had some experience with web permissions.
Finally, the least familiarity was for 3D games as only 10\% of participants played such games in web settings.

At the end of each experiment, we asked the participants to express their judgment about the game.
The 34 participants who were not tricked in all three experiments accurately realized why the button was located there and why they needed to click fast.
In the first experiment, nine participants from the tricked users group realized that they clicked on the \emph{allow} button but they were not familiar with the web permission dialog box.
As a result, they could not figure out what the consequences might be of clicking on the \emph{allow} button.
After the second experiment, eight participants who were tricked into clicking on the \emph{allow} button noticed a box appeared but could not guess what it could be.
From the ones who were tricked in the smartphone's browser, only two participants had a weak doubt whether it was a web permission dialog box or not.
Participants in the mobile browser experiment found that it was extremely difficult to avoid clicking on the \emph{allow} button while clicking rapidly.
Although the prevention technique in Google Chrome reduced the chance of fooling users by 52\%, the preventive approach cannot prevent such attacks effectively.

\subsection{Discussion}
Interestingly, among all the participants, only eight people knew how the granted permission could be revoked from the desktop and mobile browser.
This conveys that it is complicated for inexperienced users to revoke given permissions in browsers.
The findings of this study suggest that browsers should list a website's granted/blocked permissions in an easy to access manner.
In practice, browsers need to implement some preventive measures to decrease the likelihood of clickjacking attacks.
As we observed, Google chrome has considerably decreased the impact of this attack by applying a very short delay on its desktop version.
Browser vendors can employ the delay technique for the \emph{allow} button.
However, it might not be a sufficient defense mechanism against such attacks as we proved that almost half of our participants were tricked into clicking the \emph{allow} button.
To boost the prevention approach, we suggest invalidating the focus of the mouse pointer for a short period to allow users to perceive the instant UI changes.
However, finding a tradeoff between a preventive measure and user experience is a challenging problem, which requires a dedicated study. 
For instance, a long UI delay could have a negative effect on user experience.

We reported the issue to Mozilla Firefox and Google Chrome and the two companies approved the validity of the problem.
Fortunately Google Chrome in its newest version has patched this problem as follows: if the dialog box appears when the user is busy clicking, it does not accept any clicks until the user pauses for a second and clicks again.
Although the desktop version of Google Chrome is armed with the new mitigation strategy, the mobile version of this browser has no prevention mechanism yet. 
We have contacted the remaining browser vendors and are awaiting their response.

\section{Related Work}
\label{sec:related}
Rydstedt \etal analyzed the top 500 websites to study if they implemented any preventive measures regarding clickjacking attacks \cite{rydstedt2010busting}.
They found that all the defense mechanisms used by the websites can be circumvented.
To prevent clickjacking, they proposed a JavaScript-based mitigation method as a temporary measure until browsers fully support the X-FRAME-OPTIONS header.
Five new clickjacking attacks were introduced by Akhawe \etal to circumvent current UI safety specifications proposed by W3C \cite{akhawe2014clickjacking}.
They obtained a success rate between 20\% to 99\% in challenging the limitations of current defenses against UI attacks.
The proposed attacks exploit various aspects of human perception, for instance, adaptation, attention, and peripheral vision.

The ProClick framework, developed by Shahriar \etal, is designed to work as a proxy-level clickjacking detection framework \cite{shahriar2013proclick}.
Their framework works by examining the contents of requests and responses at the proxy level to detect clickjacking attacks.
Shamsi \etal developed a clickjacking prevention tool, called Clicksafe \cite{shamsi2014clicksafe}.
The developed Firefox add-on intercepts users while clicking on clickable elements with redirection code and warns users by a popup displaying necessary information.
In a study, Huang \etal proposed the InContext defense mechanism to enforce context integrity of user actions on sensitive UI elements \cite{huang2012clickjacking}.
Similarly, they developed a game in which they use a fake cursor and ask users to click rapidly on a button shown randomly in different locations in a browser.
Once the users are engaged in the game, the game switches to a Facebook \emph{Like} button at the real cursor's location, tricking the user into clicking on it.
They carried out an experiment with 2\,064 participants  and achieved a success rate ranging from 43\% to 98\% in different scenarios.
They concluded that InContext could assist participants against visual and temporal integrity clickjacking attacks.

The proposed detection tools and circumvention of defense mechanisms consider HTML elements, JavaScript code snippets, or specifications proposed by W3C. 
However, in this study, we target a browser feature, which is neither dependent on HTML elements nor JavaScript defense mechanisms.
\section{Conclusion}
\label{sec:conclusion}
We have presented Furious Clicker, a web-based game to trick users into granting web permissions.
We tested two desktop browsers and one mobile browser in order to see how users react when the permission dialog box appears.
A preliminary study showed that Furious Clicker is successful in tricking 72\% of the 120 participants.
Due to the rise of mobile phone usage nowadays, we believe that this issue has a grave impact on the security of mobile internet users.
Browsers can prevent this issue by using the delay method for buttons or by invalidating the focus of the mouse pointer to interrupt clicking.
However, finding a satisfactory tradeoff between a preventive measure and user experience is a challenging problem, which requires further study.


\section{Acknowledgments}
We gratefully acknowledge the financial support of the Swiss National Science Foundation for the project
``Agile Software Assistance'' (SNSF project No. 200020-181973, Feb. 1, 2019 - April 30, 2022).
We also thank CHOOSE, the Swiss Group for Original and Outside-the-box Software Engineering of the Swiss Informatics Society, for its financial contribution to the presentation of this paper.


\bibliographystyle{ACM-Reference-Format}
\bibliography{sample-base}

\end{document}